\documentclass[a4paper,11pt]{article}
\pdfoutput=1 % if your are submitting a pdflatex (i.e. if you have
\usepackage{jheppub} % for details on the use of the package, please
\usepackage[T1]{fontenc} % if needed

\title{\boldmath Hawking temperature and the bound on greybody factors in $D=4$ double field theory}

\author{Yang Liu}
\affiliation[a]{School of Physics and Astronomy, University of Nottingham, Nottingham NG7 2RD, UK}
\affiliation[b]{Nottingham Centre of Gravity, University of Nottingham, Nottingham NG7 2RD, UK}
\emailAdd{yang.liu@nottingham.ac.uk}
\abstract{We investigate the basic properties of Hawking radiation for spherical solutions in $D=4$ double field theory. We give the expression of the Hawking temperature for the solution and then discuss the results of various limits. We find that for all these limits only Schwarzschild solution and F-JNW solution can generate Hawking radiation. Moreover, we obtain the lower bound on greybody factors $\sigma_l(\omega)$ for the spherical solutions in $D=4$ double field theory. In particular, we calculate the bound on greybody factors $\sigma_l(\omega)$ for F-JNW solution. For F-JNW solution, $\sigma_l(\omega)$ monotonically increases with the increase of $a(b)$ for fixed $b(a)$.}

\begin{document} 
\maketitle
\flushbottom

\section{Introduction}
\label{sec:intro}
In original work in 1974 [1], Hawking showed that black holes can emit particles spontaneously at a temperature which is inversely proportional to their mass. Since then, black holes have been becoming increasingly popular in classical and quantum gravity theories [2]. Meanwhile, a series of works devoted to calculating the Hawking radiation spectrum and temperature to explore a possible theory of quantum gravity [3-5]. Here we only list some main developments in this area. For example, Page tried to calculate the particle emit rates from a nonrotating and rotating black hole in refs.[3] and [4], respectively. In 1999, in their famous paper Wilczek and Parikh [5] showed such a possibility, i.e., Hawking radiation can be considered a tunnelling process, in which particles pass through the contracting horizon of the black hole [5]. In the following years, these above results have been generalized by other researchers to other cases, such as Einstein-Gauss-Bonnet de Sitter black holes [6], charged black holes [2,7] etc.\\    
In this article, we intend to consider the properties of Hawking radiation in $D=4$ double field theory. Double field theory (DFT) is an exciting research area in string theory in recent years. The primary goal of DFT [8–11] was to reformulate supergravity with doubled coordinates, namely, $x_A = (\tilde{x}_{\mu}, x^{\nu})$, in a way that realizes T-duality as a manifest symmetry of the action and unifies diffeomorphisms and B-field gauge symmetry into ‘doubled diffeomorphisms’ [12–14]. In ref.[8], the authors have derived the most general, spherically symmetric, asymptotically flat, static vacuum solution to $D=4$ double field theory. Furthermore, in ref.[15], Stephen Angus, Kyoungho Cho and Jeong-Hyuck Park studied more general properties of Einstein double field equations. In this artilce, we intend to study some basic properties of the Hawking radiation for the solution in ref.[8]. In section 2, we review the double field theory and sperical solutions in $D=4$ double field theory briefly. In section 3, we obtain the general expression of Hawking temperature for the spherical solutions in $D=4$ double field theory. Moreover, we discuss the results of several various limits listed in ref.[8]. In section 4, we obtain the bounding of greybody factors for spherical solutions in $D=4$ double field theory, especially the case of F-JNW solution. In section 5, the results of the article have been discussed.

\section{The basics}
\subsection{Review of double field theory}
Double field theory (DFT) proposes a generalized spacetime action possessing manifest T-duality on the level of component fields [16]. Earlier efforts can be traced back to [17,18]. Due to the equivalence of winding numbers and spacetime momenta in the string spectra [16], it is natural to introduce a set of conjugated coordinates $\tilde{x}^i$, which is conjugated to winding numbers [16]. These conjugated coordinates are treated on the same footing as the usual coordinates $x_i$ [16]. The dimensionality of spacetime is doubled from $D$ to $D+D$ [16]. \\
The action of DFT unifies the metric $g_{ij}$, the two-form $b_{ij}$ and the dilaton $\phi$ by rewriting these fields in an $O(D,D,)$ covariant way, and it reduces to the supergravity action if there is no dependence on the conjugated coordinates [19]. The action is given by [19]
\begin{equation}\label{eq:2.1}
S = \int dx d\tilde{x} e^{-2d} \mathcal{R}, 
\end{equation}
where $d$ contains both the determinant of the metric and the usual dilaton $\phi$ [19], i.e.,
\begin{equation}\label{eq:2.2}
e^{-2d} = \sqrt{-g} e^{-2\phi}, 
\end{equation}
and [19,20]
\begin{equation}\label{eq:2.3}
\mathcal{R}= \frac{1}{8} \mathcal{H}^{MN} \partial_{M} \mathcal{H}^{KL} \partial_{N} \mathcal{H}_{KL} - \frac{1}{2} \mathcal{H}^{MN} \partial_{N} \mathcal{H}^{KL} \partial_{L} \mathcal{H}_{MK} - \partial_{M} d \partial_{N} \mathcal{H}^{MN} + 4 \mathcal{H}^{MN} \partial_{M} d  \partial_{N} d,
\end{equation}
where the generalized metric $\mathcal{H}_{MN}$ can be defined as [19]
\begin{equation}\label{eq:2.4}
\mathcal{H}_{MN}
=
\begin{bmatrix}
g^{ij} & -g^{ik} b_{kj}\\
b_{ik}g^{kj} & g_{ij}- b_{ik} g^{kl} b_{lj}
\end{bmatrix}.
\end{equation}
The level matching condition in closed string theory imposes the “weak constraint” $\partial \tilde{\partial} \phi (x, \tilde{x}) = 0$ for any field $\phi (x, \tilde{x})$. Furthermore, in order to ensure the action locally equivalent to the low energy effective string action, the “strong constraint” is required: $\partial \tilde{\partial} = 0$ as an operator equation, acting on any products of the fields [16]. 

\subsection{Spherical solutions in D=4 double field theory}
In this section, we will briefly review the most general form of the static, asymptotically flat and spherically symmetric vacuum solutions to $D=4$ double field theory [8,15]. Without loss of generality, we can assume that the metric for the string frame is 
\begin{equation}\label{eq:2.5}
ds^2 = e^{2\phi(r)} [-A(r)dt^2 + A(r)^{-1} dr^2 + A(r)^{-1} C(r)d\Omega^2], 
\end{equation}
where
\begin{equation}\label{eq:2.6}
 d\Omega^2= d\theta^2 + \sin^2 \theta d\phi^2.
\end{equation}
If the spacetime is asymptotically 'flat', then the metric $(2.5)$ should satisfy the following boundary conditions [8]:
\begin{equation}\label{eq:2.7}
\lim_{r \rightarrow \infty} A(r) =1,
\end{equation} 
\begin{equation}\label{eq:2.8}
\lim_{r \rightarrow \infty} r^{-2} C(r) =1,
\end{equation}
\begin{equation}\label{eq:2.9}
\lim_{r \rightarrow \infty} \phi(r) =0.
\end{equation}
From the asymptotic ‘smoothness’, the metric $(2.5)$ should satisfy [8]:
\begin{equation}\label{eq:2.10}
\lim_{r \rightarrow \infty} A'(r) = \lim_{r \rightarrow \infty} A''(r) = 0,
\end{equation} 
\begin{equation}\label{eq:2.11}
\lim_{r \rightarrow \infty} r^{-1} C'(r) = \lim_{r \rightarrow \infty} C''(r) =2,
\end{equation}
\begin{equation}\label{eq:2.12}
\lim_{r \rightarrow \infty} \phi'(r) = \lim_{r \rightarrow \infty} \phi''(r) =0.
\end{equation}
Using the form notation, the $B$-field can be written as [8]
\begin{equation}\label{eq:2.13}
B_{(2)}= \frac{1}{2} B_{\mu\nu} dx^{\mu} \wedge dx^{\nu} = B(r) \cos \theta dr \wedge d\phi + h \cos \theta dt \wedge d\phi.
\end{equation}
The $H$-flux takes the most general spherically symmetric form [8]
\begin{equation}\label{eq:2.14}
H_{(3)}= \frac{1}{3!} H_{\lambda\mu\nu} dx^{\lambda} \wedge dx^{\mu} \wedge dx^{\nu} = B(r) \sin \theta dr \wedge d\theta \wedge d\phi + h \sin \theta dt \wedge d\theta \wedge d\phi,
\end{equation}
which is closed for constant $h$ [8]. As a result, with four constants $a$, $b$, $c$, $h$ and [8]
\begin{equation}\label{eq:2.15}
c_{+} = c + \frac{1}{2} \sqrt{a^2 + b^2},
\end{equation}
\begin{equation}\label{eq:2.16}
c_{-} = c - \frac{1}{2} \sqrt{a^2 + b^2},
\end{equation}
\begin{equation}\label{eq:2.17}
\gamma_{\pm} = \frac{1}{2} (1 \pm \sqrt{1-h^2/b^2}).
\end{equation}
Then the metric $(2.5)$ can be written as [8,15]:
\begin{equation}\label{eq:2.18}
e^{2\phi} = \gamma_{+} (\frac{r- \alpha}{r+\beta})^{\frac{b}{\sqrt{a^2 + b^2}}} +  \gamma_{-} (\frac{r+ \beta}{r-\alpha})^{\frac{b}{\sqrt{a^2 + b^2}}}, 
\end{equation}
\begin{equation}\label{eq:2.19}
B_{(2)} = h \cos \theta dt \wedge d\phi, 
\end{equation}
\begin{equation}\label{eq:2.20}
H_{(3)} = h \sin \theta dt \wedge d\theta \wedge d\phi, 
\end{equation}
and
\begin{equation}\label{eq:2.21}
ds^2= e^{2\phi} [- (\frac{r- \alpha}{r+\beta})^{\frac{a}{\sqrt{a^2 + b^2}}} dt^2 +  (\frac{r+ \beta}{r-\alpha})^{\frac{a}{\sqrt{a^2 + b^2}}} \times \{dr^2 + (r-\alpha)(r+\beta) d\Omega^2\} ].
\end{equation}
If the metric must be real, then we must require $b^2 \geq h^2$ [8]. Eqs.$(2.18)$-$(2.21)$ are the most general form of the static, asymptotically flat and spherically symmetric vacuum solutions to $D=4$ double field theory [8,15]. We should point out that though the backbone of the present work is the fundamental symmetry principle of DFT, in practice, with the ansatz $(2.5)$ and $(2.13)$, we are solving the full Euler-Langrangian equations of the familiar NS-NS sector gravity, i.e. [8],
\begin{equation}\label{eq:2.22}
\int d^4 x \sqrt{-g} e^{-2\phi} (R_g + 4\partial_{\mu} \phi \partial^{\mu} \phi - \frac{1}{12} H_{\mu\nu\rho} H^{\mu\nu\rho}),
\end{equation}
as they are equivalent to the vanishing of the both two indexed and zero-indexed DFT-curvatures (i.e. DFT vacuum). The asymptotic flatness turns out to be inconsistent with the magnetic $H$-flux, and hence we put $B(r)=0$ and $H_{(3)} = h\sin \theta dt \wedge d\theta \wedge d\phi$ [8].
We can define “proper” radius [8]:
\begin{equation}\label{eq:2.23}
R \equiv \sqrt{g_{\theta \theta} (r)} = \sqrt{C(r)/A(r)} e^{\phi(r)},
\end{equation}
then the angular part of the metric can be properly normalized [8]:
\begin{equation}\label{eq:2.24}
ds^2 = g_{tt} dt^2 + g_{RR}dR^2 + R^2 d\Omega^2 = - e^{2\phi}A dt^2 + e^{2\phi} A^{-1} (\frac{dR}{dr})^{-2} dR^2 + R^2 d\Omega^2. 
\end{equation}
We list the various limits of the general solutions discussed in ref.[8], which we will use them in the later sections.\\
$\bullet$ If $b=h=0$ and $a=2M_{\infty}G > 0$, the Schwarzschild metric will be recovered: with proper radius,
\begin{equation}\label{eq:2.25}
ds^2 = -(1- \frac{2M_{\infty}G}{R})dt^2 + (1- \frac{2M_{\infty}G}{R})^{-1} dR^2 + R^2d\Omega^2, \qquad \phi=0, \qquad B_{\mu\nu}=0.
\end{equation}
If $b=h=0$ and $a=2M_{\infty}G < 0$, then
\begin{equation}\label{eq:2.26}
ds^2 = -(1+ \frac{2M_{\infty}G}{r})^{-1}dt^2 + (1+ \frac{2M_{\infty}G}{r}) dr^2 + (r+2M_{\infty}G )^2 d\Omega^2, \qquad \phi=0, \qquad B_{\mu\nu}=0.
\end{equation}
After the radial coordinate redefinition, $r \rightarrow R-2M_{\infty}G $, the metric $(2.25)$ will reduce to $(2.24)$ with negative mass [8].\\
$\bullet$ If $b=h=0$, regardless of the sign of $b$, then the metric is 
\begin{equation}\label{eq:2.27}
ds^2 = -\frac{R}{R+b}dt^2 + \frac{R}{R+b} dR^2 + R^2 d\Omega^2, \qquad e^{2\phi}=\frac{R}{R+b}, \qquad B_{\mu\nu}=0.
\end{equation}
$\bullet$ If $h=0$, the most general static spherical solution of the Einstein gravity coupled to the scalar dilaton, i.e., the F-JNW solution can be recovered [8]:
\begin{equation}\label{eq:2.28}
ds^2= -(1-\frac{\sqrt{a^2 + b^2}}{r})^{\frac{a+b}{\sqrt{a^2 + b^2}}} dt^2 + (1-\frac{\sqrt{a^2 + b^2}}{r})^{\frac{-a+b}{\sqrt{a^2 + b^2}}} [dr^2 + r(r-\sqrt{a^2+b^2})d\Omega^2],
\end{equation} 
\begin{equation}\label{eq:2.29}
e^{2\phi}= (1-\frac{\sqrt{a^2 + b^2}}{r})^{\frac{b}{\sqrt{a^2 + b^2}}}, \quad B_{\mu\nu}=0.
\end{equation}
$\bullet$ If $h=0$ and $a=b$, with the proper radius, $R=\sqrt{r^2 - \alpha^2}$, and a positive number, $\alpha \equiv \frac{1}{\sqrt{2}} |a| > 0$, the above solution reduces to [8]:
\begin{equation}\label{eq:2.30}
ds^2= - (\frac{\sqrt{R^2 + \alpha^2}-\alpha}{\sqrt{R^2 + \alpha^2}+\alpha})^{\sqrt{2}} dt^2 + \frac{R^2}{R^2+\alpha^2}dR^2 + R^2d\Omega^2, \quad e^{2\phi}= (\frac{\sqrt{R^2 + \alpha^2}-\alpha}{\sqrt{R^2 + \alpha^2}+\alpha})^{\frac{1}{\sqrt{2}}}, \quad B_{\mu\nu}=0.
\end{equation}
$\bullet$ If $h=0$ and $b^2 \geq h^2$, up to some alternative radial coordinate shift, the metric is [8]
\begin{equation}\label{eq:2.31}
ds^2= e^{2\phi} (-dt^2 + dr^2) + (r^2 + \frac{1}{4}h^2) d\Omega^2, 
\end{equation}
\begin{equation}\label{eq:2.32}
e^{2\phi} = \frac{4r^2 + h^2}{4r^2 \pm 4r \sqrt{b^2-h^2} -h^2}, \qquad H_{(3)}=h \sin\theta dt \wedge d\theta \wedge d\phi.
\end{equation}
where the sign $\pm$, coincides with that of $b$. 

\section{Hawking temperature for spherical solutions in D=4 double field theory}
In line with the analysis of refs.[5,21], we study the semiclassical tunnelling of particles through the horizon of black holes in this section. Further discussion can be seen in refs.[22-25]. On the one hand, the rate of emission $\Gamma$ will have exponential part given by [21]
\begin{equation}\label{eq:3.1}
\Gamma \sim \exp(-2ImS),
\end{equation}
where $S$ is the tunnelling action of particles and $ImS$ is the imaginary part of the tunnelling action $S$. On the other hand, according to the Planck radiation law, the emitted rate $\Gamma$ of particles with frequency $\omega$ can be written as [21]
\begin{equation}\label{eq:3.2}
\Gamma \sim \exp(-\omega/T_{BH}).
\end{equation}
Therefore, combining eqs.$(3.1)$ and $(3.2)$, the temperature at which the black hole radiates can be read off [21]:
\begin{equation}\label{eq:3.3}
T_{BH} = \frac{\omega}{2ImS}.
\end{equation}
In section 3, we will study Hawking radiation process in Painlevé-Gullstrand coordinates. Firstly, we will rewrite the metric $(2.21)$ in Painlevé-Gullstrand coordinates. We define
\begin{equation}\label{eq:3.4}
t_r = t- a(r),
\end{equation}
where $a(r)$ is a function of radial coordinate $r$. Then
\begin{equation}\label{eq:3.5}
dt^2 = dt^2_r + a'^2(r)dr^2 + 2a'dt_r dr,
\end{equation}
where $a' =da(r)/dr$. The time coordinate $t_r$, in fact, corresponds to the time measured by a stationary observer at infinity [21]. Inserting $(3.5)$ into the general static and spherical metric
\begin{equation}\label{eq:3.6}
ds^2 = -f(r)dt^2 + g(r)dr^2 + r^2 d\Omega^2,
\end{equation}
where $d\Omega^2 = d\theta^2 + \sin^2 \theta d\phi^2$, then metric $(3.6)$ can be rewritten as 
\begin{equation}\label{eq:3.7}
ds^2 = -f(r) d t^2_r - 2f(r)a' dt_r dr + \left(g(r) - f(r) a^{'2} \right) dr^2 + r^2 d \Omega^2.
\end{equation}
We require 
\begin{equation}\label{eq:3.8}
g(r) - f(r) a'^{2} =1,
\end{equation}
and choose $a' = - \sqrt{\frac{g(r)-1}{f(r)}}$. Then 
\begin{equation}\label{eq:3.9}
-2 f(r) a' = 2 \sqrt{f(r) \left(g(r) -1\right)}.
\end{equation}
Considering metric $(2.5)$, then $f(r)=e^{2\phi(r)}A(r)$ and $g(r)=e^{2\phi(r)}A^{-1}(r)$, then we have
\begin{equation}\label{eq:3.10}
-2 f(r) a' = 2 e^{2\phi(r)} \sqrt{1-A(r)},
\end{equation}
where
\begin{equation}\label{eq:3.11}
A(r)=\left(\frac{r-c_{+}}{r-c_{-}} \right)^{\frac{a}{c_{+}-c_{-}}}= \left( \frac{r-c-\frac{1}{2} \sqrt{a^2 + b^2}}{r-c+\frac{1}{2} \sqrt{a^2 + b^2}} \right)^{\frac{a}{\sqrt{a^2 + b^2}}}.
\end{equation}
Then the metric can be rewritten in Painlevé-Gullstrand coordinates as follows:
\begin{equation}\label{eq:3.12}
ds^2 = -e^{2\phi(r)}A(r)dt^2_r + 2 e^{2\phi(r)} \sqrt{1-A(r)} dt_r dr + dr^2 + e^{2\phi(r)} A^{-1}(r)C(r)d\Omega^2.
\end{equation}
In the following part of section 3, we will take $dt^2 = dt^2_r$ for convenience, i.e.,
\begin{equation}\label{eq:3.13}
ds^2 = -e^{2\phi(r)}A(r)dt^2 + 2 e^{2\phi(r)} \sqrt{1-A(r)} dt dr + dr^2 + e^{2\phi(r)} A^{-1}(r)C(r)d\Omega^2.
\end{equation}
The action for a particle moving freely in a curved background can be written as:
\begin{equation}\label{eq:3.14}
S= \int p_{\mu} dx^{\mu},
\end{equation}
with 
\begin{equation}\label{eq:3.15}
p_{\mu} = m g_{\mu\nu} \frac{dx^{\nu}}{d\sigma}, 
\end{equation}
where $\sigma$ is an affine parameter along the worldline of the particle, chosen so that $p^{\mu}$ coincides with the physical 4-momentum of the particle. For a massive particle, this requires that $d\sigma = d\tau/m$, with $\tau$ the proper time [21]. For simplicity, in this paper we only consider the case of massless scalar field $\Phi$ and ignore the angular directions. Following the spirit of ref.[21], we will obtain the formula of Hawking temrature for for spherical solutions in $D=4$ double field theory. \\
The radial dynamics of massless particles in $4d$ spacetime are determined by the equations:
\begin{equation}\label{eq:3.16}
e^{2\phi(r)}A(r)\dot{t}^2 - 2 e^{2\phi(r)} \sqrt{1-A(r)} \dot{t} \dot{r} - \dot{r}^2 = 0,
\end{equation}
\begin{equation}\label{eq:3.17}
e^{2\phi(r)}A(r)\dot{t} - e^{2\phi(r)} \sqrt{1-A(r)} \dot{r} = \omega.
\end{equation}
The second equation is the geodesic equation corresponding to the time-independence of the metric; in terms of the momentum defined in eq. $(3.15)$, it can be written $p_t = -\omega$, and so $\omega$ has the interpretation of the energy of the particle as measured at infinity [21].\\
Now we consider the the trajectory of an outgoing particle. For classical Schwarzschild solution, outgoing trajectory which crosses the horizon is forbidden [21]. The analytic continuation of an outgoing trajectory with $r > 2M$ backwards across the horizon [21], however, will give rise to an imaginary term in our action, and this term actually represents the tunnelling amplitude [5]. For an outgoing particle, eq. $(3.16)$ can be factorised to yield, the equation
\begin{equation}\label{eq:3.18}
\frac{dr}{dt} = -e^{2\phi(r)}\sqrt{1-A(r)} + e^{\phi(r)}\sqrt{e^{2\phi(r)}(1-A(r))+A(r)}.
\end{equation}
In fact, for outgoing particles, the null $\frac{dr}{dt}=0$ implies $A(r) = 0$ which is consistent with the metric $(2.5)$, i.e., at $A(r) = 0$ there exists a horizon. \\
From eq.$(3.17)$, we have
\begin{equation}\label{eq:3.19}
\frac{dr}{dt} = \frac{\dot{r}}{\dot{t}} = \frac{e^{2\phi(r)}A(r)- \frac{\omega}{\dot{t}}}{e^{2\phi(r)}\sqrt{1-A(r)}}.
\end{equation}
Combining eqs.$(3.18)$ and $(3.19)$, we obtain
\begin{equation}\label{eq:3.20}
\dot{t} = \frac{\omega}{e^{2\phi(r)}A(r) + e^{4\phi(r)}(1-A(r)) - e^{3\phi(r)} \sqrt{e^{2\phi(r)}(1-A(r))^2 + A(r)(1-A(r))}},
\end{equation}
and
\begin{equation}\label{eq:3.21}
\dot{r} = \frac{e^{2\phi(r)}A(r)\dot{t}- \omega}{e^{2\phi(r)}\sqrt{1-A(r)}}.
\end{equation}
For the case we are considering,
\begin{equation}\label{eq:3.22}
p_r = g_{rt} \dot{t} + g_{rr} \dot{r} = e^{2\phi(r)} \sqrt{1-A(r)} \dot{t} + \dot{r}.
\end{equation} 
Inserting eqs.$(3.20)$ and $(3.21)$ into $(3.22)$, we have
\begin{equation}\label{eq:3.23}
p_r = \frac{\omega \sqrt{e^{2\phi(r)}(1-A(r)) + A(r)}}{e^{\phi(r)}A(r) + e^{3\phi(r)}(1-A(r)) - e^{2\phi(r)} \sqrt{e^{2\phi(r)}(1-A(r))^2 + A(r)(1-A(r))}},
\end{equation} 
then
\begin{equation}\label{eq:3.24}
ImS= Im\int p_r dr.
\end{equation}
Then Hawking temperature is given by eq.$(3.3)$,
\begin{equation}\label{eq:3.25}
T_{BH} = \frac{\omega}{2ImS}= \frac{1}{2 Im \int \frac{\sqrt{e^{2\phi(r)}(1-A(r)) + A(r)}}{e^{\phi(r)}A(r) + e^{3\phi(r)}(1-A(r)) - e^{2\phi(r)} \sqrt{e^{2\phi(r)}(1-A(r))^2 + A(r)(1-A(r))}}dr}. 
\end{equation}
Now we will consider the various limits we have listed in section 2.2.\\
$\bullet$ If $b=h=0$, $a=2M_{\infty}G > 0$ and $\phi(r) =0$, then we have
\begin{equation}\label{eq:3.26}
p_r = \frac{\omega}{1-\sqrt{\frac{2M_{\infty}G}{r}}},
\end{equation}
then
\begin{equation}\label{eq:3.27}
T_{BH}= \frac{\omega}{2ImS} = \frac{\omega}{2Im \int \frac{\omega}{1-\sqrt{\frac{2M_{\infty}G}{r}} dr}} =\frac{1}{8\pi G M_{\infty}},
\end{equation}
which is consistent with the result of Schwarzschild black hole [21].\\
If $b=h=0$, $a=2M_{\infty}G < 0$, after the radial coordinate redefinition, $r\rightarrow R-2M_{\infty}G $, then the metric reduces to Schwarzschild metric with negative mass. Therefore, there does not exist event horizon, namely, in the usual sense the Hawking radiation does not exist for this limit.\\
$\bullet$ If $a=h=0$, regardless of the sign of $b$, from metric $(2.26)$ we have $A(r)=1$. Therefore, for this limit, the Hawking radiation does not exist.\\
$\bullet$ If $h=0$, then F-JNW solution can be recovered [8]. The location of event horizon can be obtained by setting $A(r) = 0$. From metric $(2.27)$, we have that the event horizon is located at $r_h= \sqrt{a^2 + b^2}$. Combining eqs.$(2.27)$, $(2.28)$ and $(3.25)$, we can obtain the Hawking temperature for F-JNW solution in principle although the integration is complicated.\\  
$\bullet$ If $h=0$ and $a=b$, with the proper radius, $R=\sqrt{r^2 - \alpha^2}$, and a positive number, $\alpha \equiv \frac{1}{\sqrt{2}} |a| > 0$. According to metric $(2.25)$, we can obtain the location of event horizon at $R =0$, namely, $r=\alpha \equiv \frac{1}{\sqrt{2}} |a| $. Then metric $(2.29)$ becomes $ds^2 = dr^2$. Therefore, for this limit, the Hawking radiation does not exist.\\
$\bullet$ If $a=0$ with $b^2 \geq h^2$, then from metric $(2.30)$, we obtain $A(r)=0$. In other words, no Hawking radiation exists. \\
In particular, if $a=0$ with $b^2 = h^2$, it is no longer necessary to impose the constraint $R \geq \frac{1}{2} |h|$. However, the gravity becomes repulsive and the orbital velocity becomes imaginary which has no physical sense [8].\\  

\section{The bound on greybody factors for spherical solutions in D=4 double field theory}
In section 3, we have obtained the formula of Hawking temperature for spherical solutions in $D=4$ double field theory. It is natural to discuss Hawking radiation spectrum in section 4. The Hawking radiation power $P(\omega)$ is given by [7]:
\begin{equation}\label{eq:4.1}
P(\omega) = \sum_{l} \int_{0}^{\infty} P_l(\omega) d\omega,
\end{equation} 
where
\begin{equation}\label{eq:4.2}
P_l(\omega) = \frac{A}{8\pi^2} \sigma_l(\omega) \frac{\omega^3}{\exp(\omega/T_{BH})-1},
\end{equation}
is the power emitted per unit frequency in the $l$th mode, $A$ is a a multiple of the horizon area, $l$ is the angular momentum quantum number and $\sigma_l(\omega)$ is the frequency dependent greybody factor [7].\\
Various methods have been applied to calculate greybody factors $\sigma_l(\omega)$, however, only very few cases can obtain exact analytical expressions [7]. Moreover, considering the complexity of $(3.25)$, it is difficult to obtain the general formula for Hawking radiation power $P(\omega)$. Therefore, in this section, we only briefly consider the bound on greybody factors for spherical solutions in $D=4$ double field theory following ref.[8]. Petarpa Boonserm and Matt Visser proposed that the general bounds on the greybody factor is given by [26,2,7]:
\begin{equation}\label{eq:4.3}
\sigma_l (\omega) \geq sech^2 \left(\int^{\infty}_{-\infty} \Theta dr_{\ast} \right),
\end{equation}
where
\begin{equation}\label{eq:4.4}
\Theta = \frac{\sqrt{[h'(r)^2] + [\omega^2- V_{eff} - h(r)^2]^2 }}{2h(r)}.
\end{equation}
The arbitrary function $h(r)$ has to be positive definite everywhere and satisfy the boundary condition, $h(\infty) =h(-\infty) = \omega$ for the bound eq.$(4.3)$ to hold [7] and $V_{eff}$ is the effective potential [26,2,7] which will be given later.\\
The equation of motion for a massless scalar field $\Phi$ is given by
\begin{equation}\label{eq:4.5}
\Box \Phi = 0.
\end{equation} 
Since metric $(2.5)$ has spherical symmetry, according to ref.[27], the angular variables can be seperated from the other coordinates, i.e., 
\begin{equation}\label{eq:4.6}
\Phi (t,r,\theta,\phi) = \Phi(t,r) S(\theta,\phi), 
\end{equation}
where $S(\theta,\phi)$ can be decomposed in the usual spherical harmonics satisfying [27]
\begin{equation}\label{eq:4.7}
\left(\partial^2_{\theta} + \frac{\cos \theta}{\sin \theta} \partial_{\theta} + \frac{1}{\sin^2 \theta} \partial^2_{\phi} \right) Y(\theta,\phi) = -l(l+1) Y(\theta,\phi).  
\end{equation}
In addtion, we can seperate time $t$ from from the radial coordinate, namely,
\begin{equation}\label{eq:4.8}
\Phi(t,r) = \Psi (t) \Phi(r), 
\end{equation}
where $\Psi (t) \sim e^{i \omega t}$ and satisfies $\ddot{\Psi} (t) = - \omega^2 \Psi(t)$ [27].\\
For the general static and spherical metric $(3.6)$, we can rewrite it as [27]
\begin{equation}\label{eq:4.9}
ds^2 = -f(r) (dt^2 + dr_{\ast}^2) + r^2 d\Omega^2, 
\end{equation}
by introducing the tortoise coordinate [27]:
\begin{equation}\label{eq:4.10}
r_{\ast} = \int^{r} \sqrt{\frac{g(r')}{f(r')}} dr'. 
\end{equation}
For the metric $(2.5)$, we have
\begin{equation}\label{eq:4.11}
dr_{\ast} = \frac{1}{A(r)} dr. 
\end{equation}
Then the radial equation can be writen as [2,7,26,27]:
\begin{equation}\label{eq:4.12}
\left(\frac{d^2}{d r_{\ast}^2} + \omega^2 - V_{eff} \right) u(r)=0,
\end{equation}
where 
\begin{equation}\label{eq:4.13}
V_{eff} = f(r) \frac{l(l+1)}{r^2} = e^{2\phi(r)} A(r) \frac{l(l+1)}{r^2}.
\end{equation}
Considering $(4.3)$, we obtain the bound on the greybody factor:
\begin{equation}\label{eq:4.14}
\sigma_l (\omega) \geq sech^2 \left(\int^{\infty}_{r_h} \frac{V_{eff}}{2\omega A(r)} dr \right),
\end{equation}
where $r_h$ is the location of event horizon.\\
For the various limits we considered in section 3, we found that except for Schwarzschild solution, the only solution which exists Hawking radiation is the F-JNW solution. The bound on greybody factor for Schwarzschild solution has been obtained in ref.[26]. We focus on the bound for the case of F-JNW solution.\\
Combining $(2.27)$ and $(4.14)$, we have
\begin{equation}\label{eq:4.15}
\sigma_l (\omega) \geq sech^2 \left(\frac{l(l+1)}{2\omega} \int^{\infty}_{\sqrt{a^2 + b^2}} \frac{(1-\frac{\sqrt{a^2 + b^2}}{r})^{\frac{b}{\sqrt{a^2 + b^2}}}}{r^2} dr \right).
\end{equation}  
Completing the integration, we obtain
\begin{equation}\label{eq:4.16}
\sigma_l (\omega) \geq sech^2 \left(\frac{l(l+1)}{2\omega} \frac{1}{b+\sqrt{a^2 + b^2}} \right).
\end{equation}
Figure 1 and 2 depict the lower bound of greybody factor $\sigma_l(\omega)$. From figure 1, we know that $\sigma_l(\omega)$ monotonically increases with the increase of $b$ for fixed $a$. Similarly, $\sigma_l(\omega)$ monotonically increases with the increase of $a$ for fixed $b$.
\begin{figure}[htbp]
	\centering
	\includegraphics[scale=0.4]{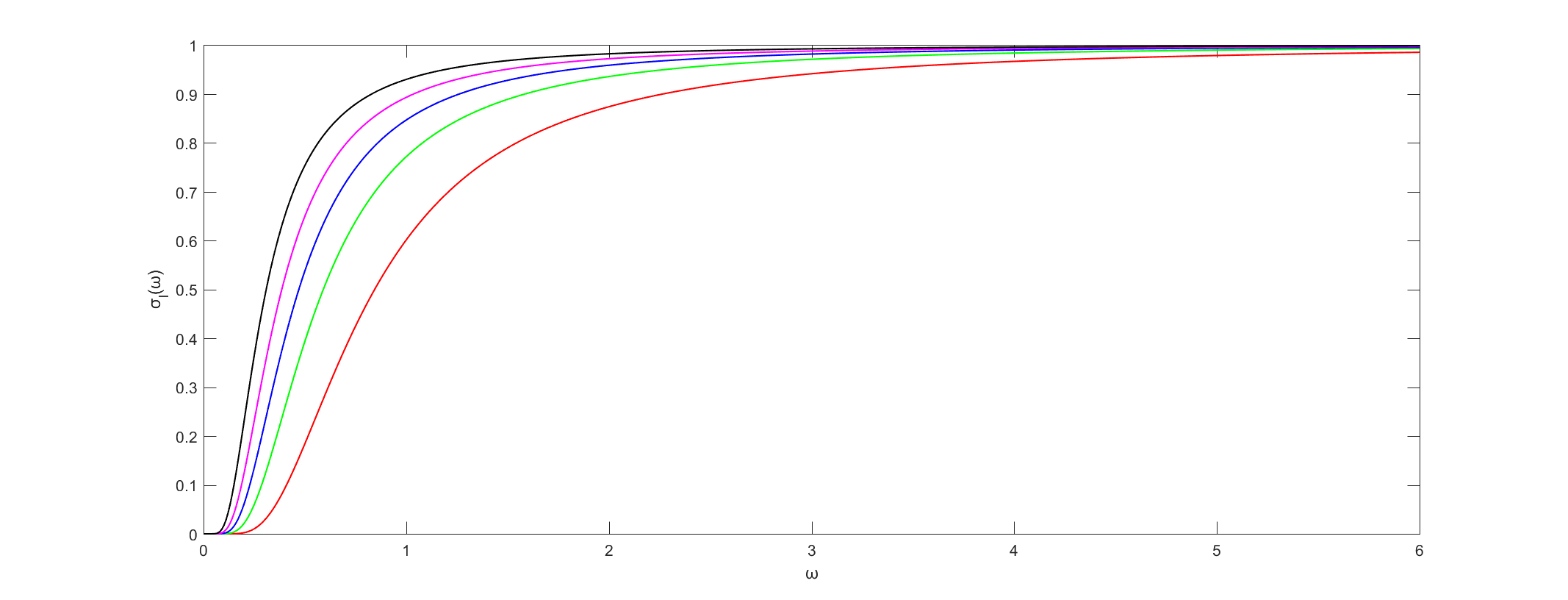}
	\caption{Lower bound of the greybody factor $\sigma_l (\omega)$ with $\omega$ for fixed value of $a$ and different values of $b$. Here we set $a=1$. The red line corresponds to $b = 0.3$, the green line to $b = 0.7$, the blue line to $b = 1$, the pink line to $b = 1.3$,  and the black line to $b = 1.7$.}
\end{figure}\\
\begin{figure}[htbp]
	\centering
	\includegraphics[scale=0.4]{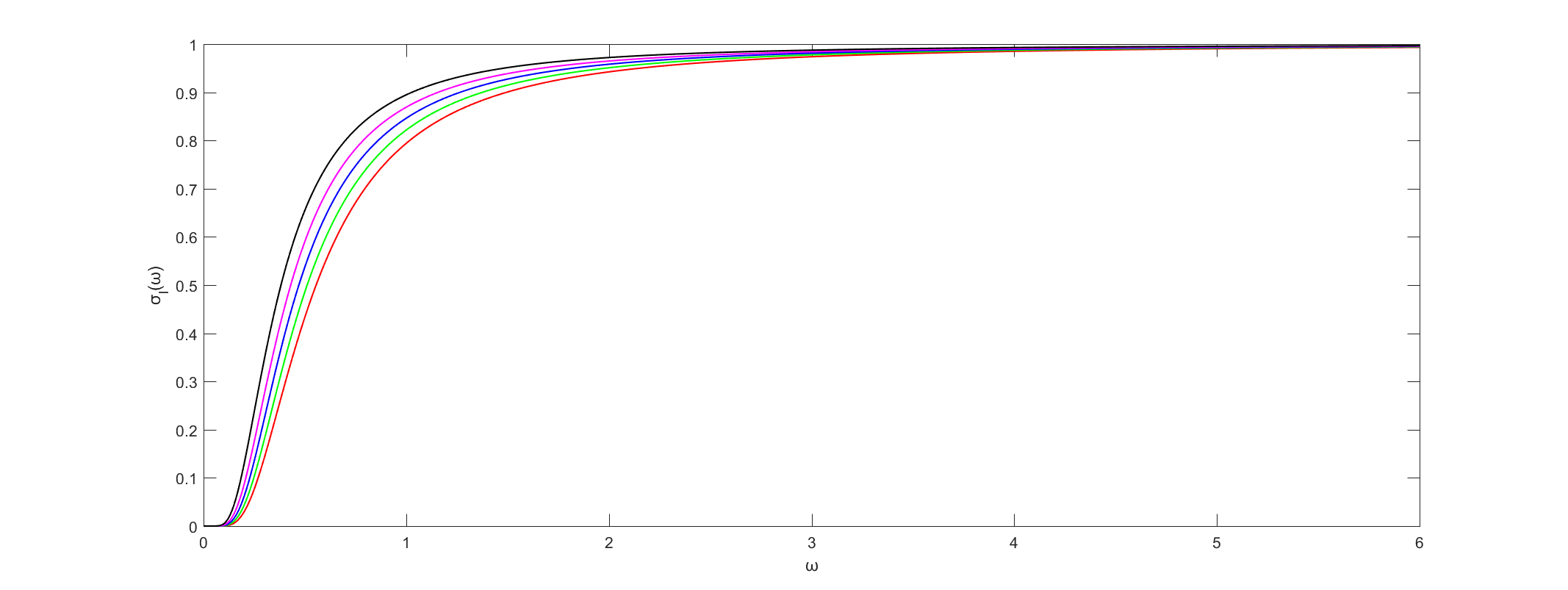}
	\caption{Lower bound of the greybody factor $\sigma_l (\omega)$ with $\omega$ for fixed value of $b$ and different values of $a$. Here we set $b=1$. The red line corresponds to $a = 0.3$, the green line to $a = 0.7$, the blue line to $a = 1$, the pink line to $a = 1.3$,  and the black line to $a = 1.7$.}
\end{figure}\\
In principle, if we combine eqs.$(2.23)$, $(2.24)$, $(3.25)$ and $(4.2)$, we can obtain the lower bound on Hawking radiation spectrum $P_l(\omega)$, but we will not do it in this article.\\

\section{Discussions and conclusions} 
On the one hand, since Hawking's original work in 1974 [1], black holes, as a fascinating and elegant object, have been increasingly popular in classical and quantum gravity theories [2]. In the following years, a series of works devoted to studying Hawking radiation, such as [3,4,5]. On the other hand, in recent years, double field theory (DFT) is an exciting research area in string theory. Sung Moon Ko, Jeong-Hyuck Park and Minwoo Suh [8] obtained the the most general, spherically symmetric, asymptotically flat, static vacuum solution to $D = 4$ double field theory. \\
In this article, we studied the basic properties of Hawking radiation in $D=4$ double field theory. Firstly, we give the formula of Hawking temperature for the sphrecial solution in $D=4$ double field theory given in ref.[8]. For all limits we considered only Schwarzschild solution and F-JNW solution can generate Hawking radiation. Furthermore, we discussed the lower bound of greybody factors $\sigma_l(\omega)$ for the spherical solution in $D=4$ double field theory. Since the bound of Schwarzshild solution has been obtained in ref.[26], we mainly focus on the F-NJW solution. For F-JNW solution, we found that $\sigma_l(\omega)$ monotonically increases with the increase of $a(b)$ for fixed $b(a)$.\\
Future work can be directed along at least three lines of further research. Firstly, the solutions of other metric in double field theory should be obtained. Secondly, in refs.[27,28] the authors have obtained quantum corrected black hole metric using effective field theory. It is natural to investigate the relationship between the two metrics in double field theory and effective field theory. Thirdly, Hawking radiation rates and information paradox in double field theory should be considered. There is therefore great potential for development of this work in the future. 


\end{document}